\documentclass[twoside,slac_one]{revtex4}
\usepackage{graphicx}
\usepackage{fancyhdr}
\usepackage{amsmath} 
\usepackage{bm}
\usepackage{amsxtra}
\usepackage{amssymb}
\usepackage{amsthm}
\usepackage{latexsym}
\usepackage{lscape}

\begin{document}

\title{Search for Supersymmetry with Photon at CMS}

%

\author{D. Nguyen for the CMS collaboration}
\affiliation{Department of Physics, Brown University, Providence, RI 02912, USA}

\begin{abstract}
We present the searches for supersymmetry (SUSY) in two channels, two photons plus missing transverse energy and a photon plus a lepton plus missing transverse energy with the CMS detector using approximately 36 $\rm{pb}^{-1}$ of pp collision at 7 TeV. No excess of events above the standard model predictions is found. Limits are set for the squark, gluino and wino masses in the general gauge-mediated SUSY context.
\end{abstract}

\maketitle

\thispagestyle{fancy}


\section{Introduction}
Supersymmetry (SUSY) is one of the most theoretical and experimental interests for physics beyond the standard model (SM). Providing the symmetry between fermionic and bosonic states, SUSY stabilizes the mass of the SM Higgs boson. The framework also provides a path to the grant unification of forces and dark matter candidates. In particular, we consider SUSY in the general gauge-mediated (GGM) breaking context~\cite{susy1,susy2,susy3,susy4,susy5,susy6,susy7}. In this scenario, SUSY is broken at energy scale much lower than the Planck scale, resulting in the gravitino ($\tilde{G}$) as the lightest SUSY particle (LSP). The next-to-lightest SUSY particles (NLSP) can be the lightest neutralino or winos. It is assumed that the neutralino and neutro winos ($\tilde{W_0}$) decays promptly to a gravitino and a photon or a Z boson while the charged wino ($\tilde{W}^{\pm}$) decays to W boson. The gravitino escapes the detection and results in missing transverse energy to the beam line. The searches for SUSY are performed at CMS in final states including two photons plus a missing transverse energy~\cite{search1} and a lepton plus a photon plus a large missing transverse energy~\cite{search2} with 36 pb$^{-1}$ at 7 TeV.
\section{Two Photons and Missing Transverse Energy}
\subsection{Selection}
The signal samples contains events with at least two photon candidates, at least one jet isolated from two photon candidates by $\Delta R = \sqrt{\Delta \eta + \Delta \phi} \geq$ 0.9 with $E_T \geq$ 30 GeV and $|\eta| \leq$ 2.6 and significant missing transverse energy. Photon are reconstructed by clustering the energy deposits in the electromagnetic calorimeter (ECAL). E$_{T}\geq$ 30 GeV and $|\eta|\leq$ 1.4 requirements are applied to the photon candidate. The energy measured in the hadronic calorimeter (HCAL) is required to be less than 5$\%$ of the ECAL energy. The isolations are used to suppress the photon background originating form the quark or gluon hadronization. These isolations are defined as the scalar sums of transverse energy of tracks or calorimeter deposition within $\Delta R=$ 0.4 of the photon direction. The energy of the candidate itself is excluded. The isolation sum for the tracker, ECAL and HCAL is required to be less than 0.001$\times E_T$ + 2.0 GeV, 0.006$\times E_T$ + 4.2 GeV and 0.0025$\times E_T$ + 2.2 GeV, respectively. $E_T$ is in GeV. We select photons which fail either the shower shape or track isolation requirement for the data-driven background estimation. These photons are referred as fake photons and most of them come from quark/gluons that give large electromagnetic deposit after hadronization. In order to distinguish photons from electrons, we search for a hit patter in the pixel detector which consistent with a track from an electron, called pixel match, and a candidate without pixel match is identified as a photon.\\
Jets are reconstructed using the anti-$k_{T}$ clustering algorithm with a size parameter of 0.5 and the jet energy is corrected using reconstructed jets. The $E_T^{miss}$ is measured in the calorimeter and corrections are applied using track momenta of charged hadrons and muons instead of calorimeter tower energies.
\subsection{Backgrounds}
The dominant backgrounds come from QCD processes such as direct diphoton, photon plus jets, and multijet production with mismeasured $E_T^{miss}$ (QCD background). A data-driven method is used to estimate these backgrounds. Two control samples, $ff$ and $ee$, are selected which are kinematically similar to the candidate sample but assumed to have no real $E_T^{miss}$. The $ff$ sample contains events with two fake photons which comes from QCD multijet process while the $ee$ sample is dominated by Z$\rightarrow$ee events with two electrons within the invariant mass window of 70 and 110 GeV. Since the hadronic energy recoils against the diphoton system, the $E_T^{miss}$ of the candidate samples can be modelled by that of control sample by reweighting the diphoton transverse energy distribution of the control sample (the weight factor is within 0.3 and 1.7) to reproduce that of candidate sample. The model $E_T^{miss}$ from the control samples are normalized with that of candidate sample in $E_T^{miss}\leq$ 20 GeV to predict the QCD backgrounds at higher $E_T^{miss}$, signal region.
Events with real $E_{T}^{miss}$, which dominated by a genuine or fake photon and a W events, are the second important background contribution. The electrons from W decays can be misidentified as photons. Therefore, this background can be estimated using $f_{e\rightarrow \gamma}$, the probability to misidentify an electron as a photon. This factor, measured with $Z\rightarrow ee$ events, is (1.4 $\pm$ 0.4)\%. We weight the $\gamma$ sample by $f_{e\rightarrow \gamma}/(1-f_{e\rightarrow \gamma})$ to estimated this electroweak background contribution.
The instrument backgrounds, which originate from the high-energy cosmic muons or beam halo with large amount of energy deposited in the ECAL, are negligible because of jet requirement.
Fig.~\ref{fig:limit_pp} (left) shows the $E_T^{miss}$ distribution in the $\gamma\gamma$ sample and the estimated backgrounds. The number of observed $\gamma\gamma$ and estimated background events in $E_T^{miss}$ are summarized in table ~\ref{closurevt}. We observe one event and the estimated total background is $1.2 \pm 0.8$. This background estimation is the average of two consistent background estimations using the $ff$ and $ee$ samples using log-normal distribution as probability density function. The common electroweak background and the correlated uncertainty due to the normalization are taken into account.
\begin{table*} [hbtp]
\caption{\label{closurevt} The event counts with $E_{T}^{miss}\geq$ 50 GeV.}
\begin{center}
\begin{tabular}{|l|c|c|c|c|}
\hline
Type &  Number of & Stat & Reweight& Normalization\\
     &   events  & error& error   &   error \\
\hline
$\gamma\gamma$ events      &1 & & & \\
Electroweak background estimate   &$0.04\pm0.03$ &$\pm0.02$&$\pm0.0$&$\pm0.01$\\

QCD background estimate (\emph{ff})         &$0.49\pm0.37$ &$\pm0.36$&$\pm0.06$&$\pm0.07$\\

QCD background estimate (\emph {ee}) &$1.67\pm0.64$
&$\pm0.46$&$\pm0.38$&$\pm0.23$\\

Total background  (using \emph {ff})  &$0.53\pm0.37$ & & &\\
Total background  (using \emph {ee})  &$1.71\pm0.64$    & & &\\
Combined total background  &$1.2\pm0.8$    & & &\\
Expected from GGM sample point  &$8.0\pm1.7$    & & &\\
\hline
\end{tabular}
\end{center}
\end{table*}
\subsection{Limits}
\begin{figure}[!Hhtb]
\centering
\includegraphics[width=0.29\textwidth]{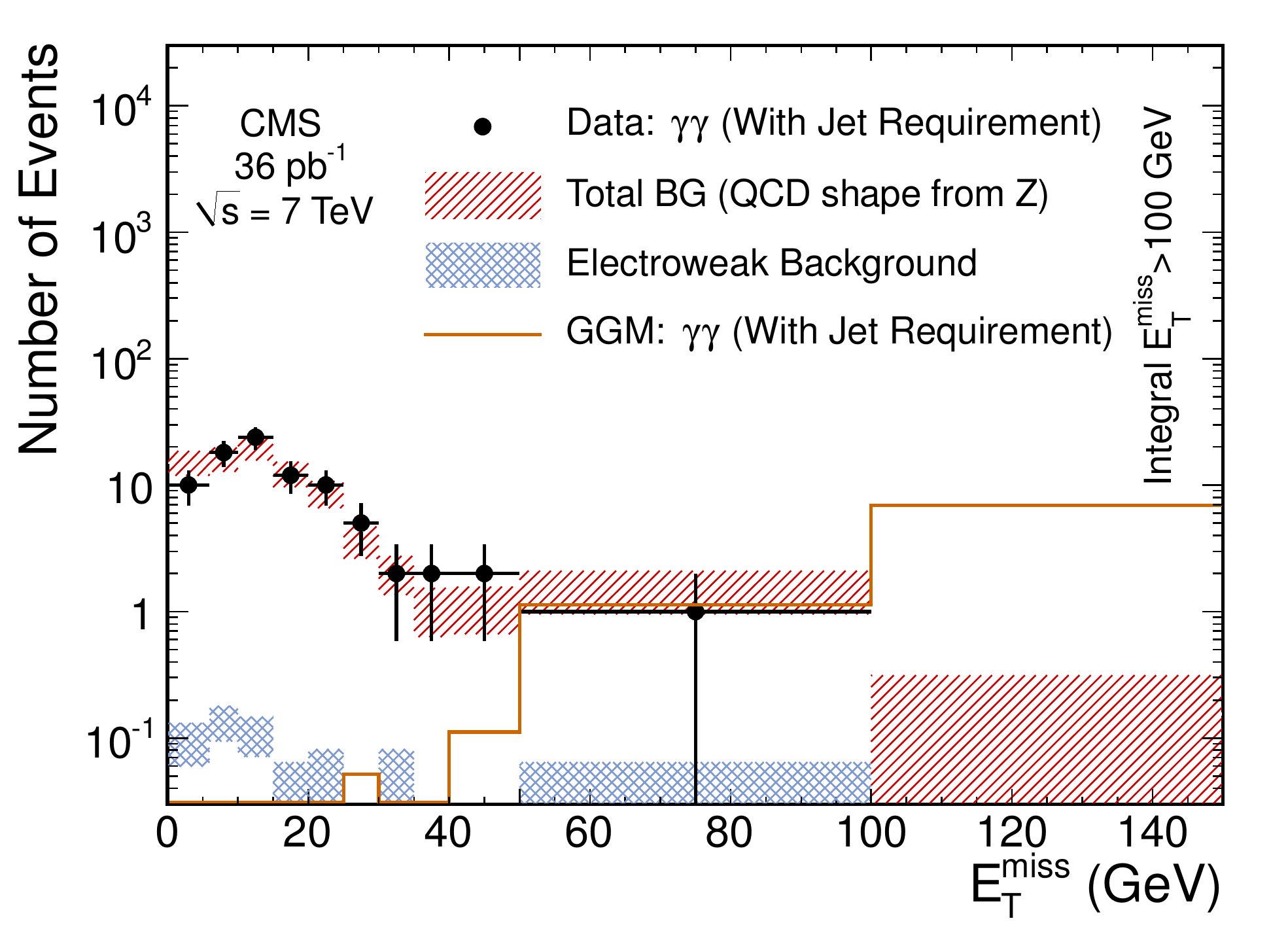}
\includegraphics[width=0.29\textwidth]{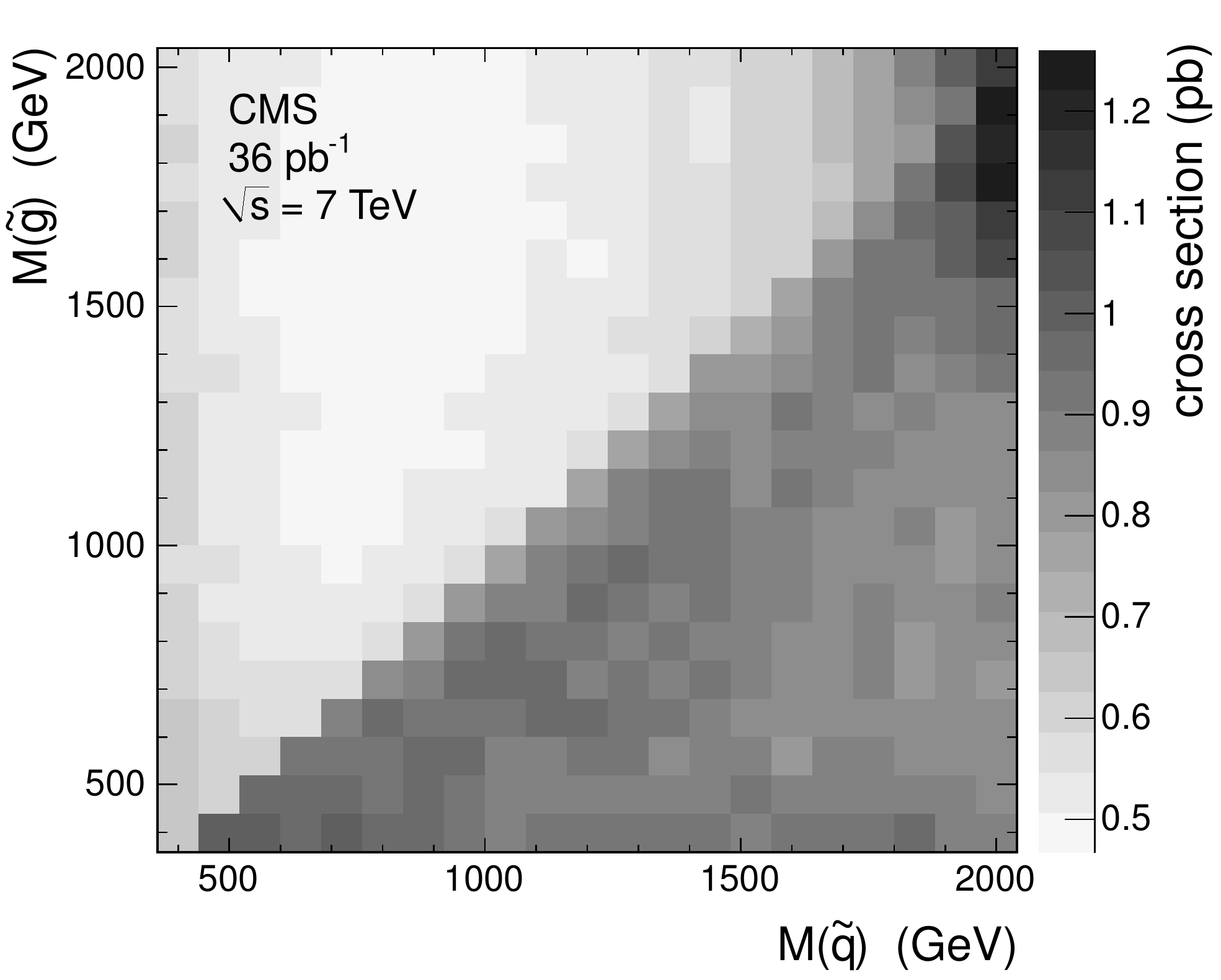}
\includegraphics[width=0.29\textwidth]{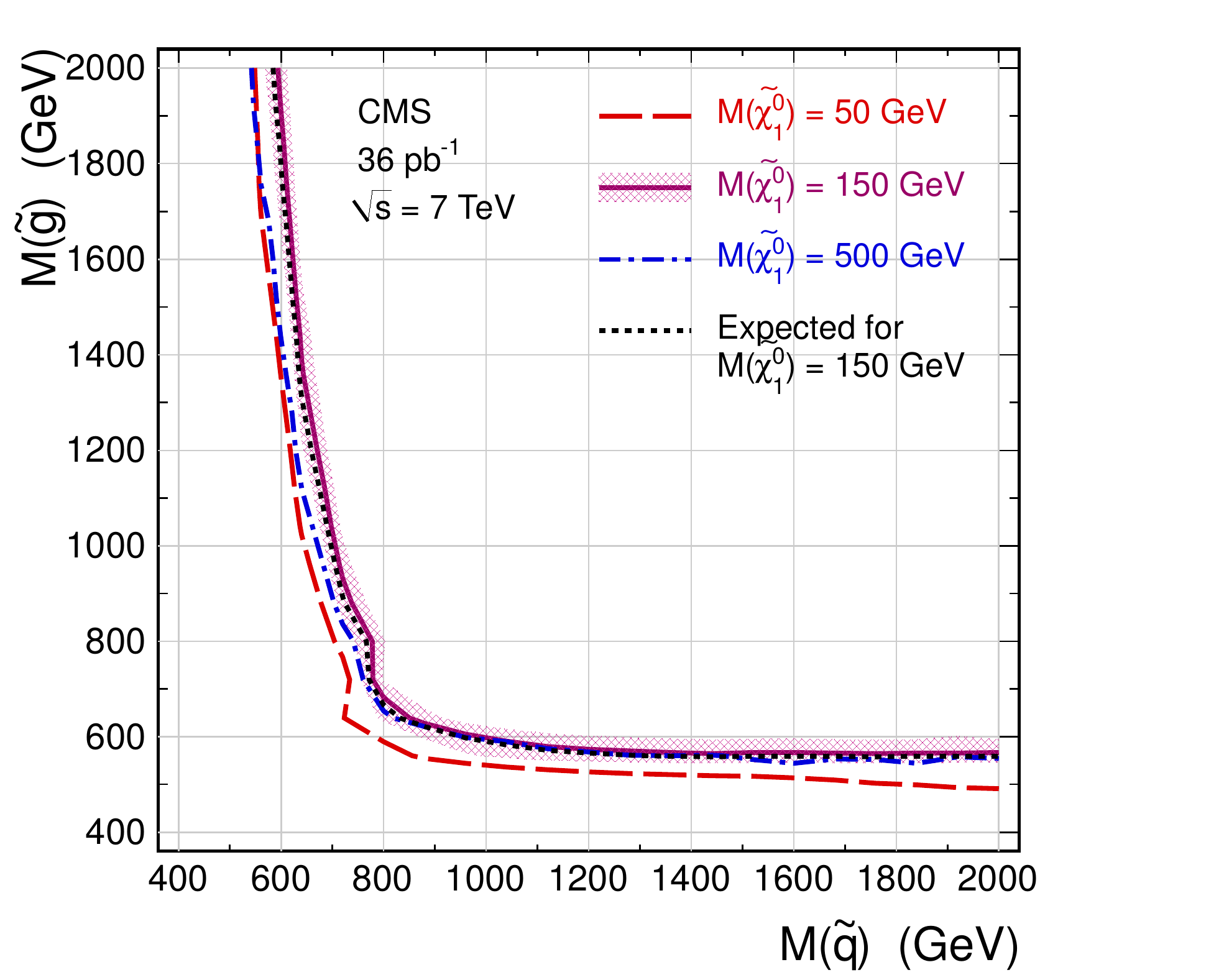}
\caption{$E_{T}^{miss}$ distribution for observed data, backgrounds and a GGM SUSY point(left). 95\% CL cross section upper limits (central). 95\% CL exclusion limits on the squark ($\tilde{q}$) and gluino ($\tilde{g}$) masses for 50, 150 and 500 GeV neutralino ($\tilde{\chi}_{1}^{0}$) masses (right).\label{fig:limit_pp}}
\end{figure}

We employ the Bayesian approach to set the upper limits on the gluino and squark production cross section. Fig.~\ref{fig:limit_pp} (central) shows the observed 95\% confident level cross section limits which vary between 0.3 and 1.1 pb for a neutralino mass of 150 GeV as a function of squark and gluino masses.\\
These cross section limits are interpreted as lower limits on squark and gluino masses using the benchmark GGM model. Fig.~\ref{fig:limit_pp} (right) shows the exclusion contours for three different choices of neutralino mass and the expected exclusion limit for a 150 GeV neutralino mass.
\section{A Lepton, a Photon and Large Missing Transverse Energy}
\subsection{Selection}
We select events with at least one high transverse momentum electron or muon and a photon. The electron candidates are reconstructed from clusters of energy deposit with a match charged tracks within ECAL barrel and endcaps. Muons are identified as charged tracks in the silicon tracking detector matched to hit patterns in the muon detectors. The leading electron or muon have to satisfy $p_{T}\geq$ 20 GeV and $|\eta|\leq$ 2.1 requirements. Electrons in the barrel-endcap transition region ($1.44<|eta|<1.57$) are excluded because of lower quality reconstructed clusters. Electron originated from a converted photon are also rejected. We further apply a set of identification and isolation criteria describe in~\cite{lpmsel}. The candidate events contain at least one photon with $p_{T}>$ 30 GeV and $|\eta|<$ 1.44 which is spatially separated from leptons by $\Delta R>$ 0.4. We require that the events contain at least one primary vertex which is within 24 cm in longitudinal direction along the beamline from the center of the detector and 2 cm in the transverse direction from the beam axis. Finally, an $E_{T}^{miss}>$ 100 GeV cut is applied.
\subsection{Background Estimation}
We estimate the dominant W$\gamma$ background using MADGRAPH MC generator~\cite{madgraph}. We correct for the next-to-leading order (NLO) precision by the a K factor from the WGRAD NLO W$\gamma$ generator~\cite{wgrad1,wgrad2}. The CTEQ6.6 NLO parton distribution function (PDF)~\cite{cteq} is used.
We use data-driven method to determine the backgrounds involving misidentified leptons and photons. Only backgrounds involved with a jet or a electron misidentified as a photon (jet$\rightarrow\gamma$ and e$\rightarrow\gamma$) are considered. The W+jet and QCD multijet production are the dominant contributions to the jet$\rightarrow\gamma$ backgrounds while the Z and $t\bar{t}$ productions dominate the e$\rightarrow\gamma$ background. The data-driven methods are based on the probability of misidentifying a jet or electron as a photon. The $l\gamma$ candidate sample in which the photon fails either the isolation or electromagnetic shower shape criterion is scaled by the jet$\rightarrow\gamma$ misidentification probability for jet$\rightarrow\gamma$ background estimation. We also scale another $l\gamma$ candidate sample in which the photon satisfies an additional requirement on the matched pattern of hits in the pixel tracker detector to estimated the e$\rightarrow\gamma$ background.\\
There are QCD contributions, which have poorly measured $E_{T}^{miss}$, to the total background. We defined a control sample which is dominated by Z$\rightarrow$ee decays with no real $E_{T}^{miss}$ and do the re-weighting to produce the $l\gamma$ sample kinematics. In particular, the $l\gamma$ transverse energy distribution is reproduced in order to ensure a correct description of the $E_{T}^{miss}$.
Table~\ref{tab_e3_met} and ~\ref{tab_mu3_met} show the background estimation results in the e$\gamma$ and $\mu\gamma$ channel, respectively. The $E_{T}^{miss}$ distribution for observed data and estimated background in which both channels are combined are shown in Fig.~\ref{fig:limit_lpm} (left).
\begin{table*}[htbp]
\begin{center}
\caption{Data and expected background event in the $e\gamma$ channel. The SUSY GMC benchmark ($m(\tilde{g})=m(\tilde{q})=450$ GeV and $m(\tilde{W}^0)=m(\tilde{W}^{\pm})=195$ GeV) signal event yields are also shown.\label{tab_e3_met}}
\begin{tabular}{|c|c|c|c|}
\hline
 & No $E_T^{miss}$ cut & $E_T^{miss} > 40$ GeV & $E_T^{miss}> 100$ GeV\\
\hline
$W\gamma$ & $44.5 \pm 9.2$     & $16.1    \pm 3.4  $  & $1.68    \pm 0.42 $ \\
jet$\rightarrow \gamma$ & $20.3 \pm 4.5$     & $3.1     \pm 0.9  $  & $0.02    \pm 0.02 $ \\
$e\rightarrow \gamma$ & $70.5 \pm  19.1$     & $0.3     \pm 0.1  $  & $0.04    \pm 0.03 $ \\
QCD                   & $134 \pm 28$    & $0.4     \pm 0.2  $  & $0.00    \pm 0.00 $ \\
\hline
Total background & $269 \pm 18$    & $19.9    \pm 3.7  $  & $1.74    \pm 0.43 $ \\
\hline
data & 264 & 16 & 1 \\
\hline
SUSY GMC prediction  & 3.94 $\pm$ 0.79 & 3.76 $\pm$ 0.75 & 2.79 $\pm$ 0.56 \\
\hline
\end{tabular}
\end{center}
\end{table*}
\begin{table}
\begin{center}
\caption{Data and expected background events in the $\mu\gamma$ channel. The SUSY benchmark GMC signal event yields are also shown. \label{tab_mu3_met}}
\begin{tabular}{|c|c|c|c|}
\hline
 & No $E_T^{miss}$ cut& $E_T^{miss} > 40 GeV$ & $E_T^{miss} > 100 GeV$ \\
\hline
$W\gamma$ & $44.8 \pm 9.3$     & $15.9    \pm 3.4  $  & $1.40    \pm 0.37 $ \\
jet$\rightarrow \gamma$ & $18.0 \pm 4.0$     & $3.7     \pm 1.1  $  & $0.10    \pm 0.09 $ \\
$e\rightarrow \gamma$ & $1.2 \pm 0.4$      & $0.6     \pm 0.2  $  & $0.09    \pm 0.04 $ \\
QCD                    & $58.3 \pm 15.1$     & $0.2     \pm 0.1  $  & $0.00    \pm 0.00 $ \\
\hline
Total background & $122.3 \pm 12.3$    & $20.4    \pm 3.7  $  & $1.59    \pm 0.39 $ \\
\hline
Data & 126 & 27 & 1 \\
\hline
SUSY GMC prediction & 5.12 $\pm$ 1.02 & 4.84 $\pm$ 0.96 & 3.66 $\pm$ 0.73 \\
\hline
\end{tabular}
\end{center}
\end{table}
\subsection{Limits}
\begin{figure}[!Hhtb]
\centering
\includegraphics[width=0.29\textwidth]{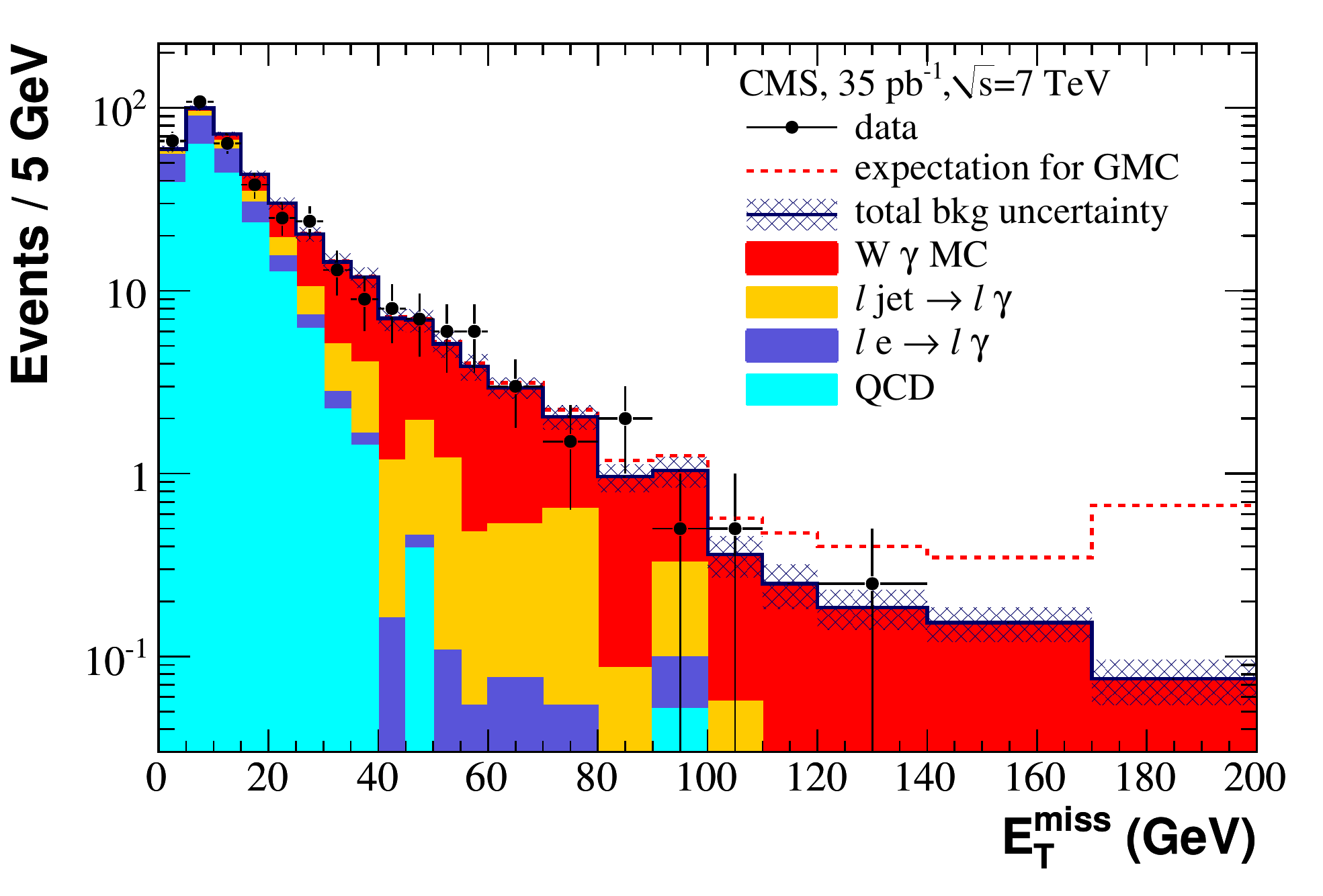}
\includegraphics[width=0.29\textwidth]{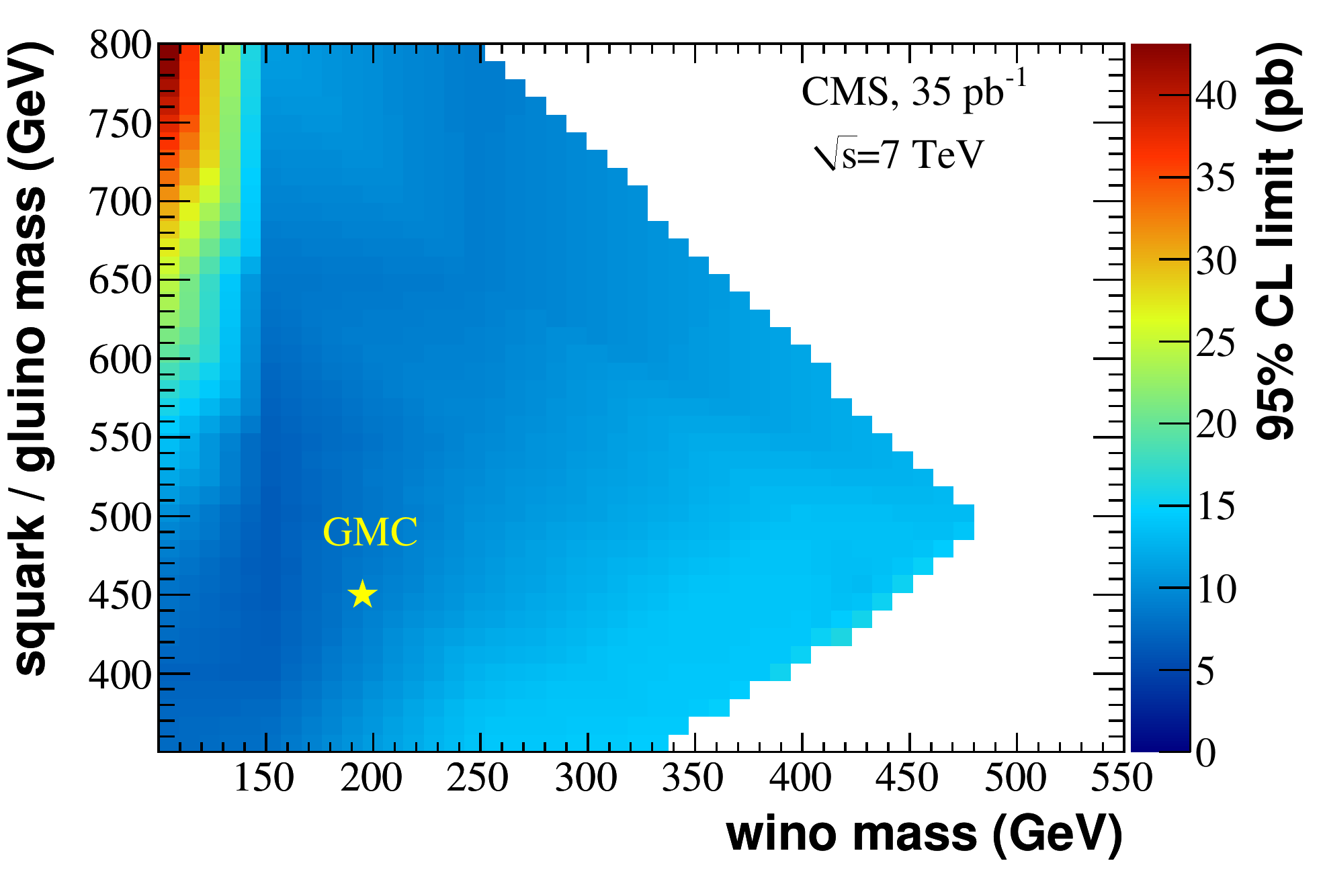}
\includegraphics[width=0.29\textwidth]{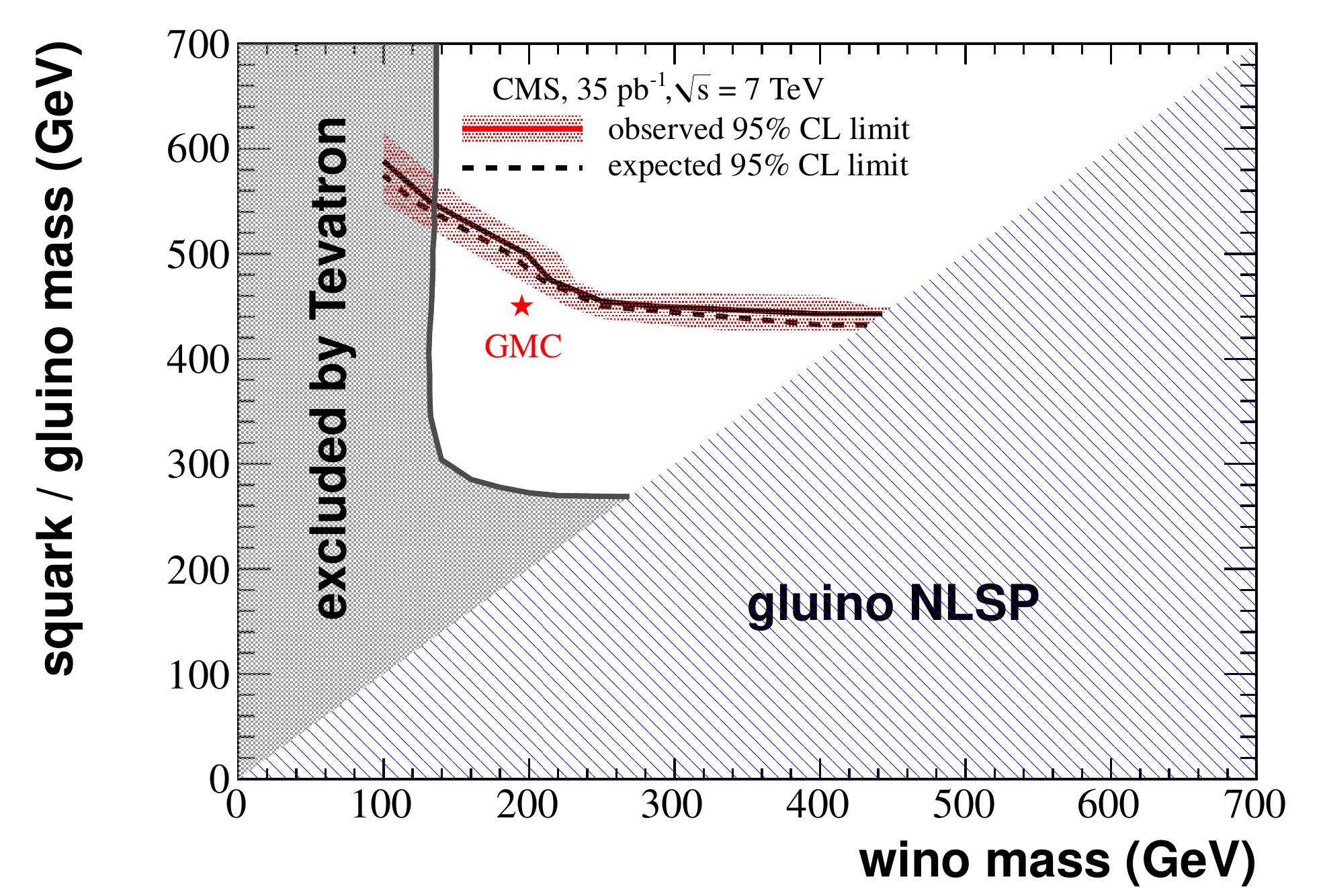}
\caption{$E_T^{miss}$ distribution for the combined e$\gamma$ andd $\mu\gamma$ samples (left). 95\% CL upper cross section limits (central). 95\% CL exclusion region as a function of the squark/gluino mass and the wino mass (right). The star indicates the GMC benchmark point.\label{fig:limit_lpm}}
\end{figure}

The Bayesian approach is used for deriving the 95\% CL upper limits on the cross sections in the SUSY model described above. The likelihoods of the e$\gamma$ and $\mu\gamma$ channels are combined in the limit calculation. Fig.~\ref{fig:limit_lpm} (central) shows these cross section limits as a function of the squark/gluino mass and the wino mass. The 95\% exclusions on the squark/gluino and wino masses are represented in Fig.~\ref{fig:limit_lpm} (right).
\section{Summary}
Searches for SUSY in the general gauge-mediated breaking models are performed either in the $\gamma +\gamma + E_T^{miss} + X$ or in the $l+\gamma + E_{T}^{miss}+X$ final states. No evidence for excess of events above the standard model expectations is found in these channels and the 95\% CL lower limits on GGM SUSY particle masses are set.


\bigskip 

\end{document}